\documentclass[amsmath,amssymb,prl,hyperlink,twocolumn,keywords]{revtex4}

	\usepackage{graphicx}
	\usepackage{soul}
	\usepackage[colorlinks=true,citecolor=blue,linkcolor=blue]{hyperref}
	\usepackage[usenames]{color}
	\usepackage{amsfonts}
	\usepackage{color}
	\usepackage{booktabs}
	\usepackage{multirow}
	\usepackage{float}
   \usepackage{titlesec}

\begin{document}

	\title{A photonic quantum interface for hybrid quantum network}
	
	\author{Jian Wang} \author{Yun-Feng Huang} \email{hyf@ustc.edu.cn}
 \author{Chao Zhang, Jin-Ming Cui, Zong-Quan Zhou, Zhi-Yuan Zhou, Jian-Shun Tang, Bi-Heng Liu}
\author{Chuan-Feng Li}
\email{cfli@ustc.edu.cn}
\author{Guang-Can Guo}
	
\affiliation{CAS Key Laboratory of Quantum Information, University of Science and Technology of China, Hefei, 230026, People's Republic of China \\Synergetic Innovation Centre in Quantum Information and Quantum Physics, University of Science and Technology of China, Hefei, Anhui 230026, China}

	
	\begin{abstract}
The hybrid quantum network, a universal form of quantum network which is aimed for quantum communication and distributed quantum computation, is that the quantum nodes in it are realized with different physical systems. This universal form of quantum network can combine the advantages and avoid the inherent defects of the different physical system. However, one obstacle standing in the way is the compatible photonic quantum interface. One possible solution is using non-degenerate, narrow-band, entangled photon pairs as the photonic interface. Here, for the first time, we generate nondegenrate narrow-band polarization-entangled photon pairs in cavity-enhanced spontaneous parametric down-conversion process. The bandwidths and central wavelengths of the signal and idler photons are 9 MHz at 935 nm and 9.5 MHz at 880 nm, which are compatible with trapped ion system and solid-state quantum memory system. The entanglement of the photon source is certified by quantum state tomography, showing a fidelity of 89.6\% between the generated quantum state with a Bell state. Besides, a strong violation against Bell inequality with $2.36\pm0.03$ further confirms the entanglement property of the photon pairs. Our method is suitable for the hybrid quantum network and will take a big step in this field.
\end{abstract}
	\maketitle

\titlespacing*{\section} {0pt}{3.5ex plus 1ex minus .2ex}{1ex plus .2ex}

Quantum network is a promising way toward the long-distance quantum communications and the large-scale distributed quantum computing \cite{1, 2}. In a quantum network, sharing entanglement among different quantum nodes is a basic requirement, and can be achieved by building quantum channels in either a deterministic \cite{3} or probabilistic manner \cite{4}. For the quantum channels, photons are a natural and excellent candidate, due to its insensitivity to environment \cite{5, 6}. Furthermore, photonic entanglement can be created in various degrees of freedom, such as polarization \cite{7}, time-bin \cite{8}, or orbital angular momentum \cite{9}. Experimental progresses have shown photonic channels can interface stationary quantum nodes into a larger quantum network \cite{2} .

Although quantum network has blossomed with single quantum systems \cite{10, 11, 12}, hybrid quantum network combining the advantages of various physical systems also begins to sprout \cite{13, 14, 15, 16}. However, there still exists a great obstacle that the flying photonic qubit should be compatible with different physical systems in a hybrid quantum network at the same time. So it's of great necessity to realize the photonic interface suitable for hybrid quantum networks. Considering the functions and aims, there are some requirements to make a well-performed photonic interface. The central frequency and bandwidth of the photons should perfectly match with the stationary physical systems in hybrid quantum networks. Last but not least, it should have the ability to entangle them in this progress. As mentioned above, entanglement is a key ingredient in any quantum networks. To satisfy these requirements, available proposals include coherent quantum-frequency conversion \cite{17}, tailoring the frequency of one kind of node to be the same as the other \cite{18}, or interconnecting the nodes with nondegenerate narrow-band photon pairs \cite{19}. Among these proposals, the first two both have some limitations: it'll be very difficult to realize the coherent quantum-frequency conversion progress if the two frequencies of photons are nearly the same, and only small amount of physical systems can be tailored with their central working frequencies. Compared with them, using non-degenerate narrow-band photon pairs as an interface seems to be the most promising one.

Narrow-band photon pair source means the bandwidth of the generated photons in frequency domain is very narrow, which matches with the atomic absorption spectrum, on the order of several MHz. For narrow-band photons, due to its long coherence time, it will show a vast advantage on the interference with independent source for long distance quantum communication on the order of several hundred Kilometres \cite{20}. Trapped atoms and ions are natural sources for narrow-band photons \cite{21, 22}. However they cannot be ideal photonic interfaces due to their complex setup and immutable working wavelengths. To be compared, cavity-enhanced spontaneous parametric down-conversion process (SPDC) is a less complicated and much more flexible method to get the narrow-band photons. In such scheme, a nonlinear crystal is inserted in the optical cavity, and the generated photons matched with the cavity mode is greatly enhanced in brightness with a narrow bandwidth depending on the cavity design. Since the first cavity-enhanced SPDC narrow band photon pair source was realized by Ou \emph{et al} \cite{23}, a lot of experiments in this area have been achieved \cite{24, 25, 26, 27, 28, 29}. Among these works, the entanglement was firstly realized in experiment by Bao \emph{et al} \cite{25} after several years' hardworking of many groups in this area, resulting in the ability of entangling the flying photons and the stationary cold atom ensemble \cite{30}. For a hybrid quantum network, the photon pair source should be nondegenerate in the frequency domain \cite{28, 29}, and it has played its role in the building of a hybrid quantum network \cite{31}. However, it has not got the ability to entangle the stationary quantum nodes into a larger quantum network, which greatly limits the application of the photonic channel to interface the different nodes in the hybrid quantum network.

 One main obstacle stands in front of the generation of polarization entanglement in nondegenerate case is the critical need of achieving four kinds of resonant modes in one optical cavity at the same time, owing to the two degrees of polarization and frequency. The best result in this area is triple resonant modes in one optical cavity so far \cite{26}. Recently, we have successfully designed a novel conjoined cavity structure, which has the potential of solving this problem \cite{29}. For entanglement generation, even with this new cavity structure, many problems and great technical challenges still remain, mainly concerned with the phase of entangled state, and the mixing of the output photons' polarization modes. As the entanglement is fragile, all these problems will kill or decrease the entanglement. So some new experimental techniques should be developed to solve these problems.

In this work, we report the generation of nondegenerate narrow-band polarization-entangled photon pair source in experiment, by solving the above problems. The central frequency and bandwidth of the photon pairs are compatible with trapped ion and rare earth-doped solid-state quantum memory systems. Trapped ion has been verified as an advanced physical system for quantum computation \cite{32} and quantum simulation \cite{33}, and rare earth-doped crystal has shown its reliable performance \cite{34} and high storage capacity \cite{15} for quantum memory. With a compatible photonic interface, this will bring us a perfect quantum network described by Kimble \cite{1}. The measured average bandwidth of generated photon pairs is 9.3 MHz. To further exploit the merit of the long coherence time of narrow-band photons, two temperature-controlled optical etalons are inserted to filter the extra multimode components.

 The experimental scheme of our photon pair source is shown in Fig. 1. It mainly consists of two parts: the generation of 453-nm pump laser and the cavity-enhanced SPDC process. The pump laser is generated from the sum frequency generation (SFG) process with a periodically poled KTiOPO4 (PPKTP) in a bow-tie cavity. The 935-nm laser (Toptica) and 880-nm laser (Ti-sapphire) here provide the power to generate the 453-nm pump laser and to lock the optical cavities in the SPDC part. It's of great necessity to get the pump laser by SFG process in this kind of experiments \cite{29}, and the SFG process in the bow-tie cavity structure can provide a highly efficient pump laser with pure optical mode. In the cavity-enhanced SPDC process part, two identically fabricated type-II PPKTPs are used to generate photon pairs. The calculated theoretical bandwidth of the phase-matching in the PPKTPs is 120 GHz. To obtain the polarization-entangled photon pairs, the method is putting the two PPKTPs side by side with optical axis perpendicular to each other, and the polarization of the pump laser is oriented with an angle of $45^{o}$ to the horizontal direction \cite{35}. Especially, in this case, the generated polarization-entangled state is:
 \begin{equation}
  |\psi\rangle=\frac{1}{\sqrt{2}}(|H_{i}\rangle|V_{s}\rangle+|V_{i}\rangle|H_{s}\rangle)
\end{equation}
where s and i represent the 935-nm signal photon and 880-nm idler photon. For cavity-enhanced SPDC process, the modes of all generated photons should match with the cavity modes. According to Eq. (1), four different modes of photons should match with the cavity modes simultaneously, which is impossible to be realized in one cavity \cite{26}. To solve this issue, a specially-designed conjoined double-cavity structure is used with the help of a customized dichroic mirror (DM) in it. Owning to the DM, the 935 nm and 880 nm down-converted photons are divided into two cavities and matched with their relative cavity independently, thus, in one cavity only two cavity modes corresponding to a single frequency need matching, which makes the experiment feasible. The optical birefringent wedges are firstly used here to compensate the nondegeneration of the polarization modes, which is simple but more compact and efficient compared with previous methods \cite{25}. Especially, it will decrease the urgent demand of the overall temperature-controlling accuracy (less than 1 mk)of the system. Owing to the fact that the phase deviation of the entangled state is greatly enhanced by the cyclical effect of the optical cavity, where the phase deviation comes mainly from tiny changes of the overall temperature and small vibrations of the system, precise temperature control of the nonlinear crystals and careful vibration isolation are persued in experiment to ensure that the phase of entangled state is stabilized in this process. Besides, in the experiment five birefringent components are inserted into the cavity to generate the entanglement, including one DM, two nonlinear crystals and two optical wedges. Due to the enhanced cyclical effect of the optical cavity, any tiny misalignments of the optic axes of these birefringent components will cause rotations of the polarization modes in the optical cavity, just working as a compound of waveplates \cite{36}. So it will cause the mixing of the output photons' polarization modes, which decreases the fidelity of the entangled state. To solve this problem, the birefringent components are mounted flexibly and aligned precisely one by one to decrease the effects. More details about the SFG progress and cavity-SPDC process are described in the supplemental material.

\begin{figure*}[tb]
   \centering
        \includegraphics[width=1\textwidth]{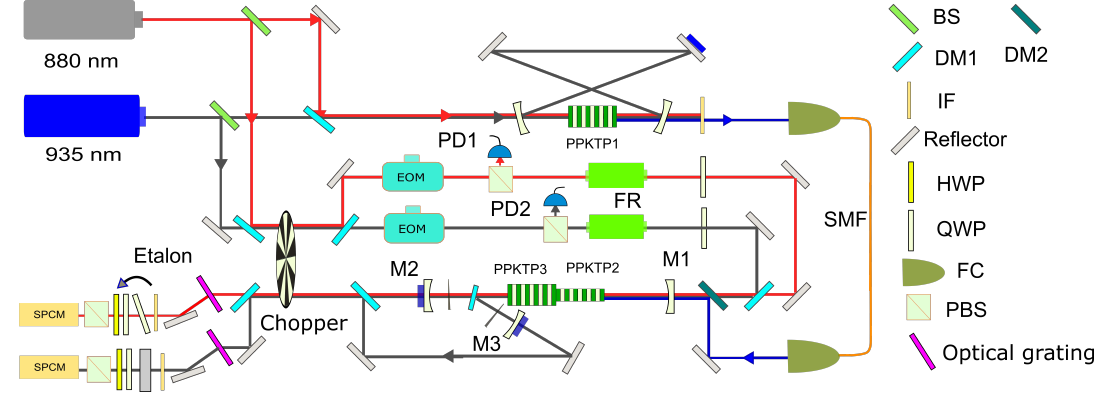}
\caption{Experimental setup. BS, beam splitter; DM, dichroic mirror; IF, interference filter; FC, fiber coupler; HWP, half-wave plate; QWP, quarter-wave plate; PBS, polarization beam splitter; EOM, electro-optic modulator; FR, farady rotator; PD1, PD2, photodiodes; SMF, single-mode fiber; SPCM, single photon counting module. The red line here means the 880-nm laser and photon, the gray line here means the 935-nm laser and photon and the blue line here means the 453-nm pump laser}
\label{1}
\end{figure*}

The cavities are locked to the reference laser beams with Pound-Drever-Hall method \cite{37}. The mechanical chopper is used to intermittently locking the cavities as before \cite{25, 28} to protect the single photon detector. In nondegenerate case, due to the inevitable difference frequency generation process in the cavity, the best method to lock the cavity is to use only one chopper. Two locking beams and SPDC photon pairs are combined together, and pass through the chopper together, which will ensure no possibilities that strong laser beams leak through the chopper to break the single photon detectors. Two temperature-controlled (10 mK) optical etalons (central wavelength 935 nm and 880 nm, bandwidth 120 MHz, FSR 8.4 GHz ) are used to filter out the extra multimode components in the cavity outputs.

\begin{figure}[tb]
    \centering
        \includegraphics[width=0.48\textwidth]{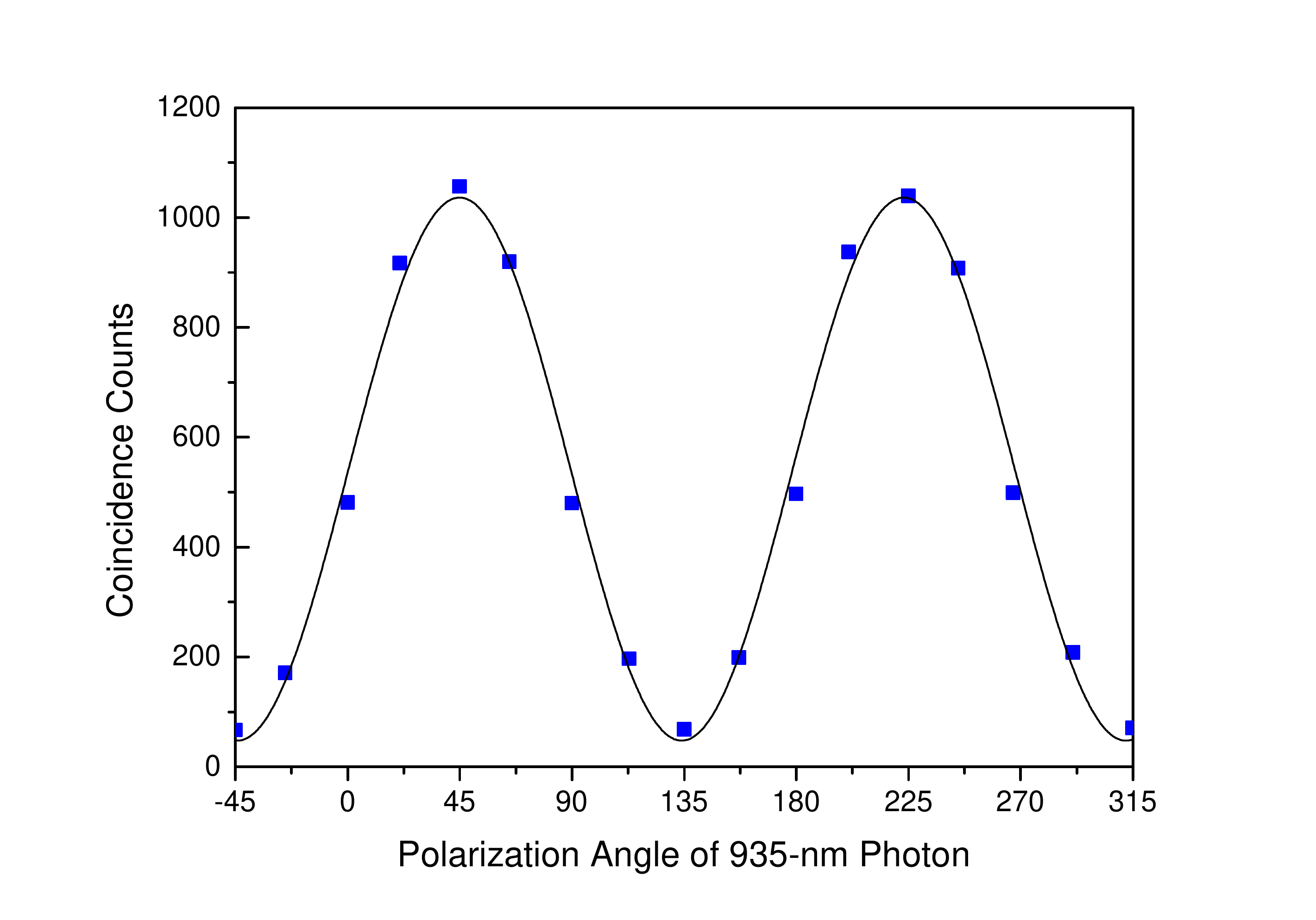}
\caption{The polarization correlation for the entangled photon source. The coincidence time window here is 80 ns. The pump power is 9 mW and the integration time is 200 s. The angle in horizontal axis here has the unit in degree.
}
\label{fig2}
\end{figure}

To verify the entanglement property of the photon source, we first measure the polarization correlations with adjustable polarization analyzers, each consisting of a HWP, a QWP, and a PBS. The analyzer of 880 nm is set to $-45^{o}$, while the analyzer of 935 nm is varied by rotating its HWP. The result is shown in Fig. 2. We also conduct the Clauster-Horne-Shimony-Holt (CHSH) inequality test and got $S=2.36\pm0.03$, which means that the violation is 12 standard deviations \cite{38}. To get the full characterization of the generated entangled photon source, we also perform the quantum state tomography, and reconstruct the density matrix, which is shown in Fig, 3. From the tomography result, we obtain a state fidelity of 89.6\% between the generated state and the targeted Bell state $|\psi\rangle$  \cite{39}.

\begin{figure}[tb]
    \centering
        \includegraphics[width=0.50\textwidth]{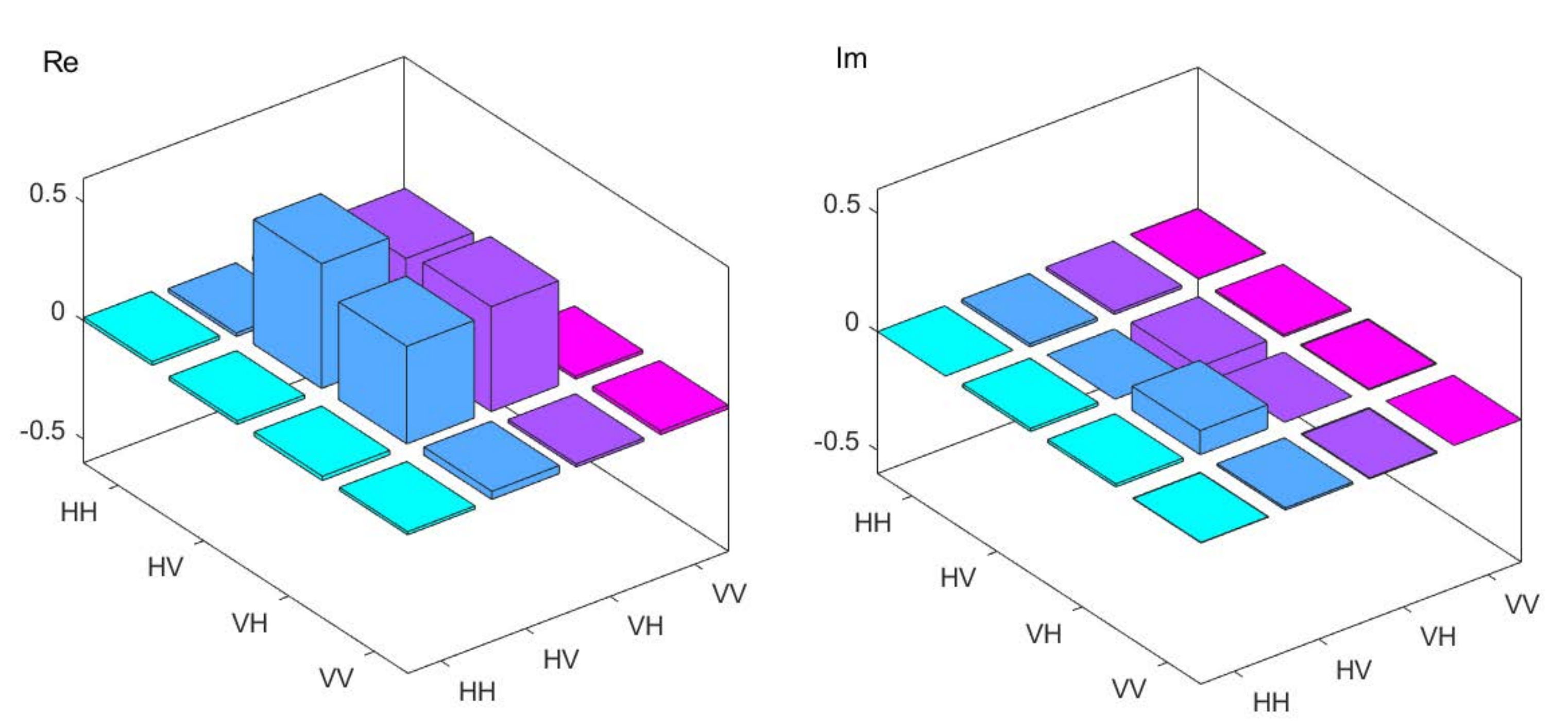}
        \caption{Entangled state tomography results. The left is the real part, and the right is the imaginary part.
}
\label{fig4}
\end{figure}

 \begin{figure}[tb]
    \centering
        \includegraphics[width=0.50\textwidth]{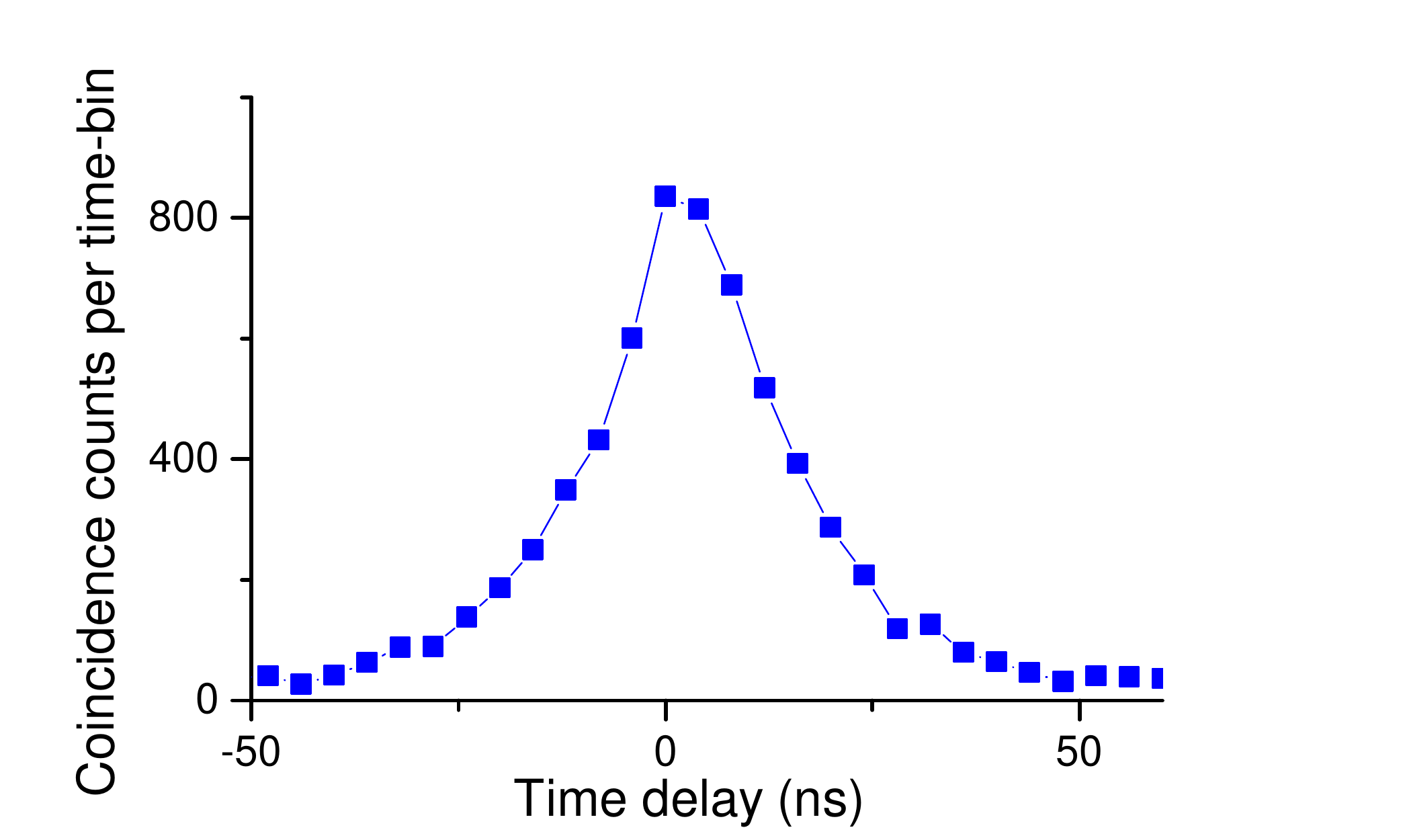}
\caption{The correlation function $G_{S,I}^{(2)}(\tau)$ is measured at a pump power of 9 mW. The time-bin size here is 4 ns, and the integration time is 1800 s.
}
\label{fig3}
\end{figure}

 \begin{figure}[tb]
    \centering
        \includegraphics[width=0.50\textwidth]{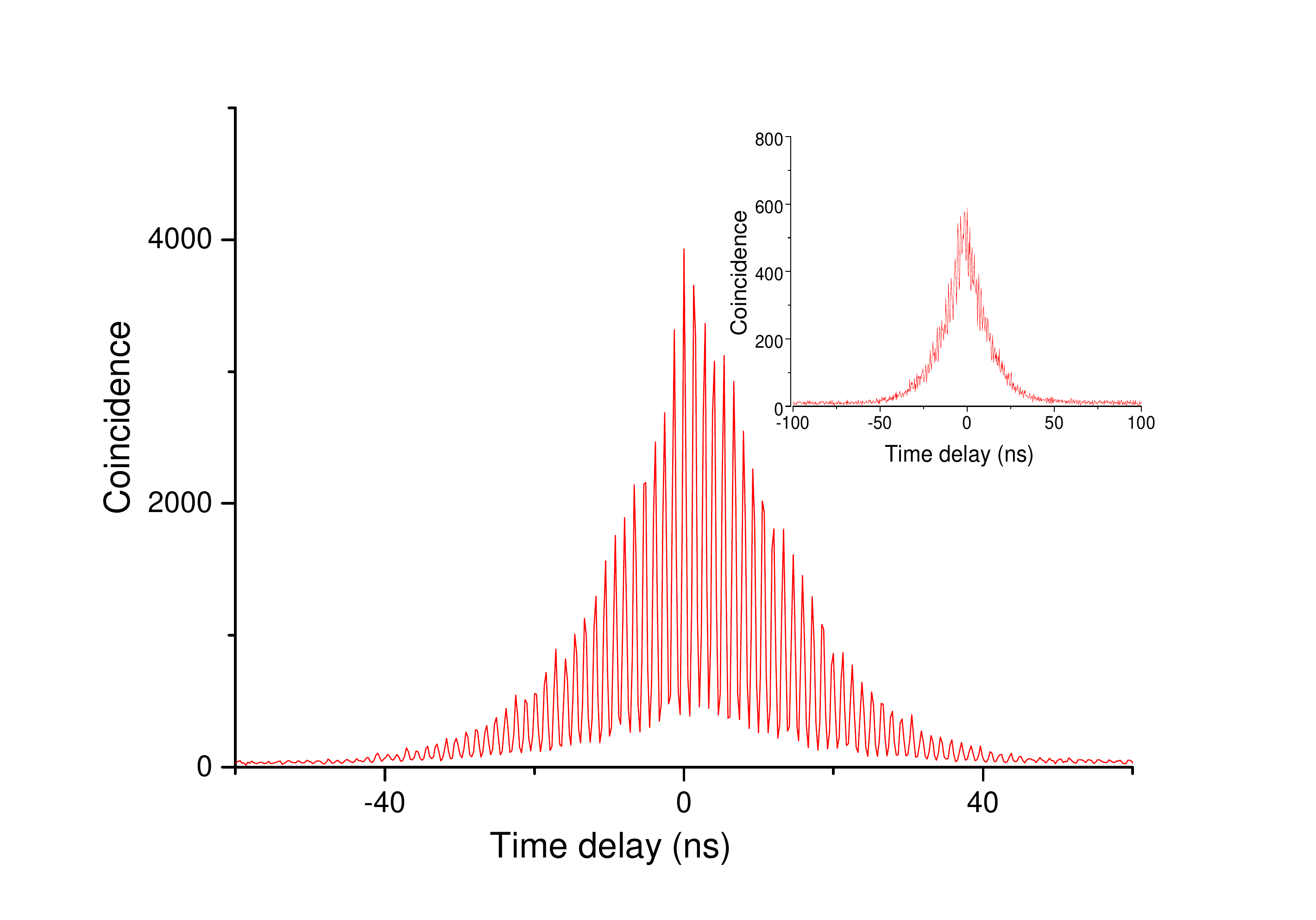}
\caption{The time-resolved measurement of correlation function $G_{S,I}^{(2)}(\tau)$ is performed at a pump power of 9 mW. The time-bin size here is 256 ps, and the integration time is 600 s. The inset is measured with etalons used to filter the multimode components. The pump power here is also 9 mW and the integration time is 3600 s
}
\label{fig5}
\end{figure}

 The bandwidth of the photon source is inversely proportional to the correlation time between the generated signal and idler photons. So we conduct the time-correlated measurements, specifically to measure the second-order cross-correlation function, $G_{S,I}^{(2)}(\tau)$, between the signal and idler photons  \cite{28}. The time distribution of photon pairs arriving at the detectors is recorded by the time-to-digital-converter (Picoquant 400). The measured result is  plotted in Fig. 4. We use the fitting function $e^{-2\pi\Delta\upsilon\tau}$ to fit on the two sides of the curve separately, and get the signal and idler photons bandwidths are 9 MHz at 935 nm and 9.5 MHz at 880 nm. The tiny difference should be blamed for the different losses of the inserted optical devices at the two wavelengths.

Owing to property of optical cavity, the generated cavity-enhanced narrow band photon pair often has comb-like multimode structure in frequency domain, which will extremely decrease the coherence of the source \cite{25, 28}. In our experiment, we insert two FP etalons into the two outports of the photon source to filter the extra multimode components. To compare the unfiltered and filtered cases, we take the time-resolved measurement of the cross-correlated function $G_{S,I}^{(2)}(\tau)$ on the output photon pairs in both cases. For the existence of multimode photons, the time-resolved measurement of $G_{S,I}^{(2)}(\tau)$ will give out a time-domain comb-like structure of the curve, resulted from the interference between different frequency modes. On the contrary, the time-domain comb-like structure will disappear under the situation of efficient elimination of multimode components. The results are shown in Fig. 5. The relatively smooth curve can imply that the multimode components in the source are efficiently filtered.

The rate of single mode photon pair source is about $5\textrm{/s}$ at pump power of $9$ mW. So the normalized spectrum brightness is approximately $0.062\textrm{/s/MHz/mW}$ without any modification. And it has a fidelity of 89.6\% between the generated quantum state with a Bell state. However, these values have big potentials to be better with this method. The low counting rate is mainly caused by the intra-cavity loss and the low detector efficiencies. For the fidelity of the photon pair source, there's a strong demand on the stabilization of phase in the entangled state and precise alignment of optical axis of the five birefringent devices inserted. Besides, we also find the strong noise in the PPKTP crystals, which decreases the ratio of signal to noise and gives a limitation of the 453-nm pump power. All these described above can be improved by using much more better equipments. More details can be obtained in the supplementary material.

In conclusion, we have realized a photonic channel suitable for hybrid quantum network through the way of generating nondegenerate narrow-band entangled photon source. It satisfies the needs for photonic interface because it can not only match the central frequencies and bandwidths of different kinds of quantum nodes well, but also entangle them. The generation of this kind of source takes a big step toward hybrid quantum network.

\vspace{0.1cm}
 This work was supported by the National Natural Science Foundation of China (Nos. 61327901, 61490711, 11325419, 61225025, 11474268, 11374288, 11304305, 11404319), the Strategic Priority Research Program (B) of the Chinese Academy of Sciences (Grant No. XDB01030300), the National Program for Support of Topnotch Young Professionals (Grant No. BB2470000005), the Fundamental Research Funds for the Central Universities (WK2470000018)£¬Key Research Program of Frontier Sciences, CAS (Grant No. QYZDY-SSW-SLH003).


\vspace{0.6cm}

\begin{thebibliography}{10}
	\providecommand{\url}[1]{\texttt{#1}}
	\providecommand{\urlprefix}{URL }
	\providecommand{\eprint}[2][]{\url{#2}}
    \bibitem{1} J. Kimble, Nature (London) \textbf{453}, 1023 (2008).
	
	\bibitem{2} L.-M. Duan and C. Monroe, Rev. Mod. Phys. \textbf{82}, 1209 (2010).

    \bibitem{3}  S. Ritter, C. N\"{o}lleke, C. Hahn, A. Reiserer, A. Neuzner, M. Uphoff, M. M\"{u}cke, E. Figueroa, J. Bochmann, and G. Rempe, Nature (London) \textbf{484}, 195 (2012).


    \bibitem{4}  D. L. Moehring, P. Maunz, S. Olmschenk, K. C. Younge, D. N. Matsukevich, L.-M. Duan, and C. Monroe, Nature (London) \textbf{449}, 68 (2007).

	
	\bibitem{5} B. Hensen,	H. Bernien,	A. E. Dr\"{e}au, A. Reiserer, N. Kalb, M. S. Blok,	J. Ruitenberg,	R. F. L. Vermeulen,	R. N. Schouten,	C. Abell\"{a}n,	W. Amaya,	V. Pruneri,	M. W. Mitchell,	M. Markham, D. J. Twitchen,	D. Elkouss,	S. Wehner,	T. H. Taminiau and R. Hanson, Nature (London) \textbf{526}, 682 (2015).

	\bibitem{6} X.-S Ma, T. Herbst,	T. Scheidl,	D. Wang, S. Kropatschek, W. Naylor,	B. Wittmann, A. Mech, J. Kofler,	E. Anisimova, V. Makarov, T. Jennewein,	R. Ursin and A. Zeilinger, Nature (London) \textbf{489}, 269 (2012).

    \bibitem{7} Y.-F. Huang, B.-H. Liu, L. Peng, Y.-H. Li, L. Li, C.-F. Li,and G.-C. Guo, Nature Communication. \textbf{2}, 546 (2011).
	

	\bibitem{8} N. Sangouard, C. Simon, H. de Riedmatten, and N. Gisin, Rev. Mod. Phys. \textbf{83}, 33 (2011).
	
	\bibitem{9} A. Mair, A. Vaziri, G. Weihs, A. Zeilinger, Nature (London) \textbf{6844}, 313 (2001)

     \bibitem{10} C. Monroe, R. Raussendorf, A. Ruthven, K. R. Brown, P. Maunz, L.-M. Duan, and J. Kim, Phys. Rev. A \textbf{89}, 022317 (2014).
	
     \bibitem{11} Z.-S. Yuan, Y.-A. Chen, B. Zhao, S. Chen, J. Schmiedmayer, and J.-W. Pan, Nature \textbf{454}, 1098 (2008)

     \bibitem{12} A. Delteil, Z. Sun, W.-b. Gao, E. Togan, S. Faelt and A. Imamo\v{g}lu, Nature Physics \textbf{12}, 218 (2016)
	
	
	\bibitem{13} M. Wallquist, K. Hammerer, P. Rabl, M. Lukin and P. Zoller, Phys. Scr. \textbf{T137}, 014001 (2009).
	
	\bibitem{14} H. M. Meyer, R. Stockill, M. Steiner, C. Le Gall, C. Matthiesen, E. Clarke, A. Ludwig, J. Reichel, M. Atat¨¹re, and M. K\"{o}hl, Phys. Rev. Lett. \textbf{114}, 123001 (2015).
	
	\bibitem{15} J.-S. Tang, Z.-Q. Zhou, Y.-T. Wang, Y.-L. Li,	X. Liu,	Y.-L. Hua,	Y. Zou,	S. Wang, D.-Y. He, G. Chen, Y.-N. Sun, Y. Yu,	M.-F. Li, G.-W. Zha, H.-Q. Ni, Z.-C. Niu, C.-F. Li and G.-C. Guo, Nature Communication. \textbf{6}, 8652 (2015).
	
	\bibitem{16} R. W. Andrews,	R. W. Peterson,	T. P. Purdy, K. Cicak, R. W. Simmonds, C. A. Regal and  K. W. Lehnert, Nature Physics \textbf{10}, 321 (2014).

    \bibitem{17} H. M. Meyer, R. Stockill, M. Steiner, C. Le Gall, C.Matthiesen, E. Clarke, A. Ludwig, J. Reichel, M. Atatre, and M. K\"{o}hl, Phys. Rev. Lett. \textbf{114}, 123001 (2015).

    \bibitem{18} S. Ramelow, A. Fedrizzi, A. Poppe, N. K. Langford, and A. Zeilinger, Phys. Rev. A \textbf{85}, 013845 (2012).

	\bibitem{19} C. Simon, H. de Riedmatten, M. Afzelius, N. Sangouard, H. Zbinden, and N. Gisin, Phys. Rev. Lett. \textbf{98}, 190503 (2007).

    \bibitem{20} R. Kaltenbaek, B. Blauensteiner, M. \"{z}ukowski, M. Aspelmeyer, and A. Zeilinger, Phys. Rev. Lett. 96, 240502 (2006).

	\bibitem{21} J. K. Thompson, J. Simon, H. Loh, and V. Vuletic, Science \textbf{313}, 74 (2006).
	
	\bibitem{20}  M. Schug, J. Huwer, C. Kurz, P. M\"{u}ller, and J. Eschner, Phys. Rev. Lett. \textbf{110}, 213603 (2013).
		
	\bibitem{23} Z. Y. Ou and Y. J. Lu, Phys. Rev. Lett. \textbf{83}, 2556 (1999).

	
    \bibitem{24}  C. E. Kuklewicz, F. N. C. Wong, and J. H. Shapiro, Phys.Rev. Lett. \textbf{97}, 223601 (2006).
	
    \bibitem{25} X.-H. Bao, Y. Qian, J. Yang, H. Zhang, Z.-B. Chen, T. Yang, and J.-W. Pan, Phys. Rev. Lett. \textbf{101}, 190501 (2008).

	\bibitem{26} M. Scholz, L. Koch, and O. Benson, Phys. Rev. Lett. 102,063603 (2009).

    \bibitem{27} F. Wolfgramm, Y. A. de Icaza Astiz, F. A. Beduini, A.Cer\`{e}, and M. W. Mitchell, Phys. Rev. Lett. \textbf{106}, 053602 (2011).

	\bibitem{28} J. Fekete, D. Riel\"{a}nder, M. Cristiani, and H. de Riedmatten, Phys. Rev. Lett. \textbf{110}, 220502 (2013).

      \bibitem{29} J. Wang, P.-YJ Lv, J.-M. Cui, B.-H. Liu, J.-S. Tang, Y.-F. Huang, C.-F. Li, and G.-C. Guo, Phys. Rev. Applied \textbf{4}, 064011 (2015).

    \bibitem{30} H. Zhang, X.-M. Jin, J. Yang, H.-N. Dai, S.-J. Yang, T.-M. Zhao, J. Rui, Y. He, X. Jiang, F. Yang,	G.-S. Pan, Z.-S. Yuan, Y.-j. Deng, Z.-B. Chen, X.-H. Bao, S. Chen, B. Zhao and J.-W. Pan, Nature Photonics \textbf{5}, 628 (2011).

	\bibitem{31} D. Riel\"{a}nder, K. Kutluer, P. M. Ledingham, M. G\"{u}ndo\u{g}an, J. Fekete, M. Mazzera, and H. de Riedmatten, Phys. Rev. Lett. \textbf{112}, 040504 (2014).

	
		
	\bibitem{32} T. Monz, P. Schindler, J. T. Barreiro, M. Chwalla, D. Nigg, W. A. Coish, M. Harlander, W. H\"{a}nsel, M. Hennrich,and R. Blatt, 14-Qubit Entanglement: Creation and Coherence, Phys. Rev. Lett. \textbf{106}, 130506 (2011).
	
    \bibitem{33} J.-M. Cui, Y.-F. Huang, Z. Wang, D.-Y. Cao, J. Wang, W.-M. Lv, Y. Lu, L. Luo, A. del Campo, Y.-J. Han, C.-F. Li, G.-C. Guo, Sci. Rep. \textbf{6}, 33381 (2016).
	
    \bibitem{34} Z.-Q. Zhou, W.-B. Lin, M. Yang, C.-F. Li, and G.-C. Guo, Phys. Rev. Lett. \textbf{108}, 190505 (2012).

	\bibitem{35} R. Rangarajan, M. Goggin, and P. Kwiat, Optics Express \textbf{17}, 018920 (2009).

    \bibitem{36} J. Yang, T.-M. Zhao, H. Zhang, T. Yang, X.-H. Bao, and J.-W. Pan, Chin. Phys. B. \textbf{20}, 024202, (2011)
	
	\bibitem{37}  E. D. Black, Am. J. Phys. \textbf{69}, 79 (2001).

    \bibitem{38}  X.-L. Niu, Yun.-Feng Huang, G.-Y Xiang,G.-C Guo, and Z.-Y Ou, Optics Letters \textbf{33}, 9 (2008).

    \bibitem{39}  A. G. White, D. F. V. James, P. H. Eberhard, and P. G.Kwiat, Phys. Rev. Lett. \textbf{83}, 3103 (1999).



	
\end{thebibliography}
\end{document}